\documentclass[12pt,a4paper]{article}
\usepackage{feynmp}
\begin{document}
\begin{fmffile}{procpics}
\title{Towards a Complete Calculation of $\gamma\gamma\to4f$%
  \thanks{Contribution to the proceedings of "The International
      Europhysics Conference on High-Energy Physics", 19-26 August
      1997, Jerusalem, Israel.}
  \thanks{Supported by Deutsche Forschungsgemeinschaft.}
  \thanks{Presented by E.\,B., supported by the HEP97 Organizing
      Committee and Russian Ministry of Science.}}
\author{%
  E.\,Boos${}^{1,2}$\thanks{e-mail: \texttt{boos@ifh.de}},
  T.\,Ohl${}^2$\thanks{e-mail: \texttt{Thorsten.Ohl@Physik.TH-Darmstadt.de}}\\
  ${}^1$Moscow State University, Russia\\
  ${}^2$Darmstadt University of Technology, Germany}
\maketitle
\begin{abstract}
  We present a general classification of all four fermion final states
  in $\gamma\gamma$~collisions and a calculation of cross sections
  below the $W^+W^-$-threshold.
\end{abstract}
The planned $e^+e^-$~Linear Collider will provide the opportunity to
study $\gamma\gamma$-collisions with energies of several hundred GeV
in the center of mass system.  Of immediate interest will be gauge and
Higgs boson production.  However, neither the Higgs, nor the~$W$ or
the~$Z$ will be directly observed and we typically have to study four
fermion production processes, for which the resonant diagrams
$\gamma\gamma\to VV\to4f$ do \emph{not} form gauge invariant subsets
and calculations of~$\gamma\gamma\to4f$ beyond on-shell gauge boson
production are required for precision studies.

As a step towards the construction of a complete event generator for
$\gamma\gamma\to4f$, we have introduced~\cite{Boos/Ohl:1997} a
classification of all four fermion final states in
$\gamma\gamma$-collisions and the corresponding Feynman diagrams.
\begin{figure}
  \begin{center}
    \mbox{($Q$)
    \fmfset{arrow_angle}{5}
    \fmfset{arrow_len}{2mm}
    \begin{fmfgraph*}(25,15)
      \fmfleft{g,g'}
      \fmflabel{$\gamma$}{g}
      \fmflabel{$\gamma$}{g'}
      \fmfright{f1,f2b,f3,f4b}
      \fmflabel{$f_1$}{f1}
      \fmflabel{$\bar f_2$}{f2b}
      \fmflabel{$f_3$}{f3}
      \fmflabel{$\bar f_4$}{f4b}
      \fmf{photon,t=2}{g,qgv,g'}
      \fmf{boson,lab=\tiny$W$,lab.side=left}{v1,qgv,v2}
      \fmf{fermion}{f2b,v1,f1}
      \fmf{fermion}{f4b,v2,f3}
    \end{fmfgraph*}
    \qquad
    ($D$)
    \begin{fmfgraph*}(25,15)
      \fmfleft{g,g'}
      \fmflabel{$\gamma$}{g}
      \fmflabel{$\gamma$}{g'}
      \fmfright{f1,f2b,f3,f4b}
      \fmflabel{$f_1$}{f1}
      \fmflabel{$\bar f_2$}{f2b}
      \fmflabel{$f_3$}{f3}
      \fmflabel{$\bar f_4$}{f4b}
      \fmf{photon,t=2}{g,tgv1}
      \fmf{photon,t=2}{tgv2,g'}
      \fmf{boson,lab=\tiny$W$,lab.side=left}{v1,tgv1,tgv2,v2}
      \fmf{fermion}{f2b,v1,f1}
      \fmf{fermion}{f4b,v2,f3}
    \end{fmfgraph*}
    \qquad
    ($T$)
    \begin{fmfgraph*}(25,15)
      \fmfleft{g,g'}
      \fmflabel{$\gamma$}{g}
      \fmflabel{$\gamma$}{g'}
      \fmfright{f1,f2b,f3,f4b}
      \fmflabel{$f_1$}{f1}
      \fmflabel{$\bar f_2$}{f2b}
      \fmflabel{$f_3$}{f3}
      \fmflabel{$\bar f_4$}{f4b}
      \fmf{photon,t=2}{g,v1}
      \fmf{photon,t=2}{tgv,g'}
      \fmffixedx{0}{v1,v2}
      \fmf{boson,lab=\tiny$W$,lab.side=left}{v2,tgv,v3}
      \fmf{fermion}{f2b,v2,v1,f1}
      \fmf{fermion}{f4b,v3,f3}
    \end{fmfgraph*}}\\[3\baselineskip]
    \mbox{($S$)
    \begin{fmfgraph*}(25,15)
      \fmfleft{g,g'}
      \fmflabel{$\gamma$}{g}
      \fmflabel{$\gamma$}{g'}
      \fmfright{f1,f4b,f3,f2b}
      \fmflabel{$f_1$}{f1}
      \fmflabel{$\bar f_2$}{f2b}
      \fmflabel{$f_3$}{f3}
      \fmflabel{$\bar f_4$}{f4b}
      \fmf{photon,t=2}{g,v1}
      \fmf{photon,t=2}{g',v3}
      \fmf{boson,lab=\tiny$W,,Z,,\gamma,,g$,tension=0.2}{v2,v4}
      \fmf{fermion,tension=0.2}{f4b,v4,f3}
      \fmf{fermion}{f2b,v3,v2,v1,f1}
    \end{fmfgraph*}
    \qquad
    ($B$)
    \begin{fmfgraph*}(25,15)
      \fmfleft{g,g'}
      \fmflabel{$\gamma$}{g}
      \fmflabel{$\gamma$}{g'}
      \fmfright{f1,f2b,f3,f4b}
      \fmflabel{$f_1$}{f1}
      \fmflabel{$\bar f_2$}{f2b}
      \fmflabel{$f_3$}{f3}
      \fmflabel{$\bar f_4$}{f4b}
      \fmf{photon,t=2}{g,v1}
      \fmf{photon,t=2}{g',v2}
      \fmffixedx{0}{v1,v2}
      \fmf{boson,lab=\tiny$W,,Z,,\gamma,,g$,lab.side=left}{v3,v4}
      \fmf{fermion}{f2b,v3,v2,v1,f1}
      \fmf{phantom}{v3,v2}
      \fmf{fermion}{f4b,v4,f3}
    \end{fmfgraph*}
    \qquad
    ($M$)
    \begin{fmfgraph*}(25,15)
      \fmfleft{g,g'}
      \fmflabel{$\gamma$}{g}
      \fmflabel{$\gamma$}{g'}
      \fmfright{f1,f2b,f3,f4b}
      \fmflabel{$f_1$}{f1}
      \fmflabel{$\bar f_2$}{f2b}
      \fmflabel{$f_3$}{f3}
      \fmflabel{$\bar f_4$}{f4b}
      \fmf{photon,t=2}{g,v1}
      \fmf{photon,t=2}{g',v4}
      \fmf{boson,lab=\tiny$W,,Z,,\gamma,,g$,lab.side=left,tension=0.5}{v2,v3}
      \fmf{fermion}{f2b,v2,v1,f1}
      \fmf{fermion}{f4b,v4,v3,f3}
    \end{fmfgraph*}}
  \end{center}
  \caption{\label{fig:Q/D/T/S/B/M}%
    The six topologies for~$\gamma\gamma\to4f$ in unitary gauge
    (see~\cite{Boos/Ohl:1997} for $R_\xi$-gauges).}
\end{figure}
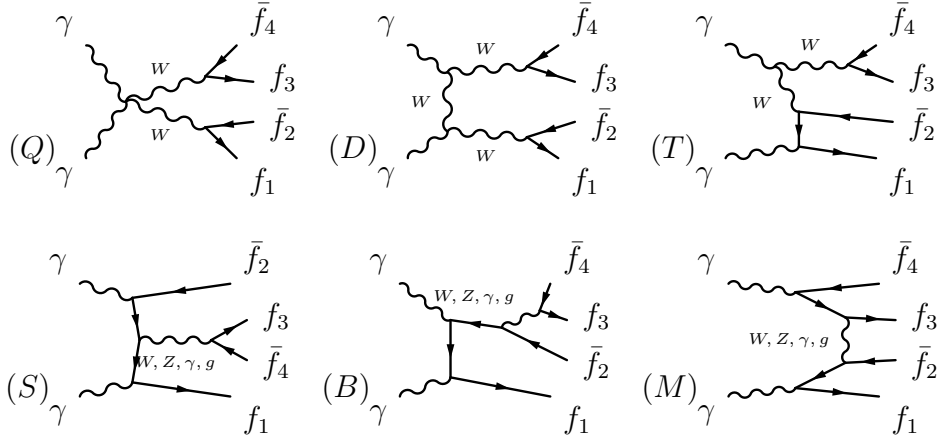
Fig.~\ref{fig:Q/D/T/S/B/M} shows all six topologies appearing in the
Feynman diagrams for $\gamma\gamma\to4f$.  The flavors of the final
state fermions determine which of~$W$, $Z$, $\gamma$, or~$g$
contribute for the lines labelled such.
In a three generation standard model, there are 98 possible four
fermion final states that fall into only eight different classes.
They are listed in table~\ref{tab:classes} with the number of diagrams
for each topology.
\begin{table}
  \begin{center}
    \begin{tabular}{lc|rrrrrr|r|c|c}
      \multicolumn{2}{c|}{Class}
                             &$Q$& $T$& $D$& $S$& $B$& $M$& $\sum$
                     & $\mathcal{O}(\alpha_S)$
                             & e.\,g. \\\hline
      \textit{CC13}  & l      & 1 &  4 &  2 &  0 &  4 &  2 &     13
                     &       & $e^-\bar\nu_e \mu^+\nu_\mu$ \\
      \textit{CC21}  & sl     & 1 &  6 &  2 &  2 &  6 &  4 &     21
                     &       & $e^-\bar\nu_e u\bar d$ \\
      \textit{CC31}  & h      & 1 &  8 &  2 &  4 &  8 &  8 &     31
                     &       & $u\bar d\bar cs$ \\
      \textit{NC06}  & l/sl   & 0 &  0 &  0 &  2 &  4 &  0 &      6
                     &       & $\nu_e\bar\nu_e \mu^-\mu^+$ \\
      \textit{NC20}  & l/h    & 0 &  0 &  0 &  4 &  8 &  8 &     20
                     & $+10$ & $e^-e^+e^-e^+$ \\
      \textit{NC40}  & l/sl/h & 0 &  0 &  0 &  8 & 16 & 16 &     40
                     & $+20$ & $e^-e^+\mu^-\mu^+$ \\
      \textit{mix19} & l      & 1 &  4 &  2 &  2 &  8 &  2 &     19
                     &       & $e^-e^+\nu_e\bar\nu_e$ \\
      \textit{mix71} & h      & 1 &  8 &  2 & 12 & 24 & 24 &     71
                     & $+20$ & $u\bar u d\bar d$
    \end{tabular}
  \end{center}
  \caption{\label{tab:classes}%
    The eight classes of diagrams in $\gamma\gamma\to 4f$.}
\end{table}
The charged current classes (\textit{CC}$n$) are gauge invariant
completions of $W$-pair production and, unlike~$e^+e^-\to4f$, there
are \emph{no} final states without multi-peripheral diagrams.  The
neutral current classes (\textit{NC}$n$) do not contain any cubic or
quartic gauge couplings.  Finally, the mixed classes (\textit{mix}$n$)
correspond to final states that are invariant under the transformation
from ``charge exchange'' to ``charge retention'' form and vice versa.
For each class, one typical final state is given
(see~\cite{Boos/Ohl:1997} for complete lists).  The
column~$\mathcal{O}(\alpha_S)$ shows the number of one gluon exchange
diagrams to be added in the neutral current classes for purely
hadronic final states.
\begin{figure}
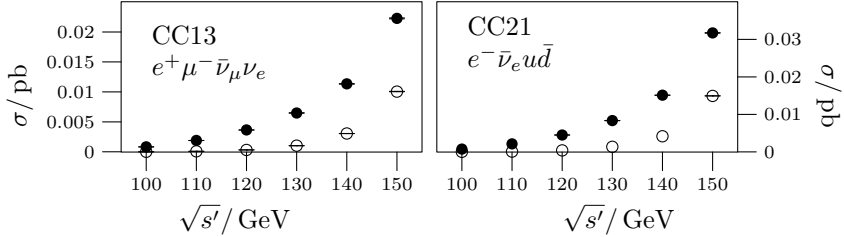

  \begin{center}
    \mbox{\includegraphics{procplots.1}\includegraphics{procplots.2}}
  \end{center}
  \caption{\label{fig:plots}%
    Integrated cross sections for charged current final states below
    the $W^+W^-$-threshold (full circles: full calculation,
    open circles: $\gamma\gamma\to WW^*$).}
\end{figure}

%%%%%%%%%%%%%%%%%%%%%%%%%%%%%%%%%%%%%%%%%%%%%%%%%%%%%%%%%%%%%%%%%%%%%%%%

\end{fmffile}
\end{document}